\documentclass{IEEEtran}
\usepackage{cite}
\usepackage{amsmath,amssymb,amsfonts}

\usepackage{graphicx}
\usepackage{textcomp}
\usepackage{amsmath,amsfonts}
\usepackage{algorithm}
\usepackage{algpseudocode}
\usepackage{tabularx}
\usepackage{subfigure}
\usepackage{subcaption}
\usepackage{booktabs}
\usepackage{siunitx}
\usepackage{placeins}

\usepackage{setspace}
\def\BibTeX{{\rmB\kern-.05em{\sci\kern-.025emb}\kern-.08emT\kern-.1667em\lower.7ex\hbox{E}\kern-.125emX}}
    
\begin{document}
\title{A 2-bit Ku-band Digital Metasurface with \\Infinitely Scalable Capability}
\author{Xiaocun Zong, \IEEEmembership{Student Member, IEEE}, Hao Shi, Fan Yang*, \IEEEmembership{Fellow, IEEE}, Yong Liu, \\Shenheng Xu, \IEEEmembership{Member, IEEE}, Maokun Li, \IEEEmembership{Fellow, IEEE}
\thanks{This work is supported by the National Key Research and Development
Program of China under Grant No. 2023YFB3811501. (Corresponding Author: \textit{Fan Yang})}
\thanks{
The authors are with the Department of Electronic Engineering, State Key laboratory of Space Network and Communications, Tsinghua University, Beijing 100084, China.
}

}
\maketitle

\begin{abstract}
In this letter, we present the design and implementation of a 2-bit digaital metasurface operating in the Ku-band, engineered to exhibit advanced polarization conversion characteristics and support dual-polarization control for both X- and Y-polarizations. To address the challenge of array size scalability hindered by extensive DC control routing in 2-bit metasurfaces, we propose a novel RF-DC separation architecture. This approach integrates the metasurface and DC control circuitry onto separate printed circuit boards (PCBs), interconnected via pin cascading, enabling theoretically unlimited two-dimensional array expansion. To validate this design, a ${4\times16 \times 16}$ metasurface prototype was fabricated and experimentally evaluated, which can achieve a gain of 28.3dB and an aperture efficiency of 21.02\%, confirming the scalability and performance of the proposed architecture. The developed 2-bit high-gain metasurface offers significant reference value for applications in long-distance communication and radar detection. Furthermore, the RF-DC separation architecture introduces a pioneering framework for large-scale metasurface deployment in practical engineering scenarios, enhancing design flexibility and scalability.
\end{abstract}

\begin{IEEEkeywords}
Metasurface, 2-bit, polarization conversion, beam scanning, RF-DC separation, array scalability.
\end{IEEEkeywords}

\section{Introduction}
\IEEEPARstart{M}{etasurfaces}, also known as reconfigurable intelligent surfaces (RIS), have emerged as a transformative technology in electromagnetic wave manipulation, offering unprecedented control over amplitude, phase, and polarization of electromagnetic waves \cite{b1,b2,b3,b4, cui2014coding, liaskos2018new}. These planar, subwavelength-structured arrays enable dynamic beamforming, polarization conversion, and wavefront shaping, making them highly valuable for applications such as wireless communications, radar systems, and imaging \cite{zhang2020reconfigurable, b12,b17,b20}. By integrating tunable elements, such as diodes or varactors, metasurfaces can dynamically reconfigure their electromagnetic properties, providing a compact and energy-efficient alternative to traditional phased arrays and reflectarrays. The advent of metasurfaces has thus opened new avenues for enhancing system performance in terms of gain, directivity, and adaptability, particularly in the microwave and millimeter-wave frequency bands\cite{yang2016programmable, b9,b10,b11,b13}.
\begin{figure}[t]
	\centering
	\includegraphics[width=0.55\linewidth]{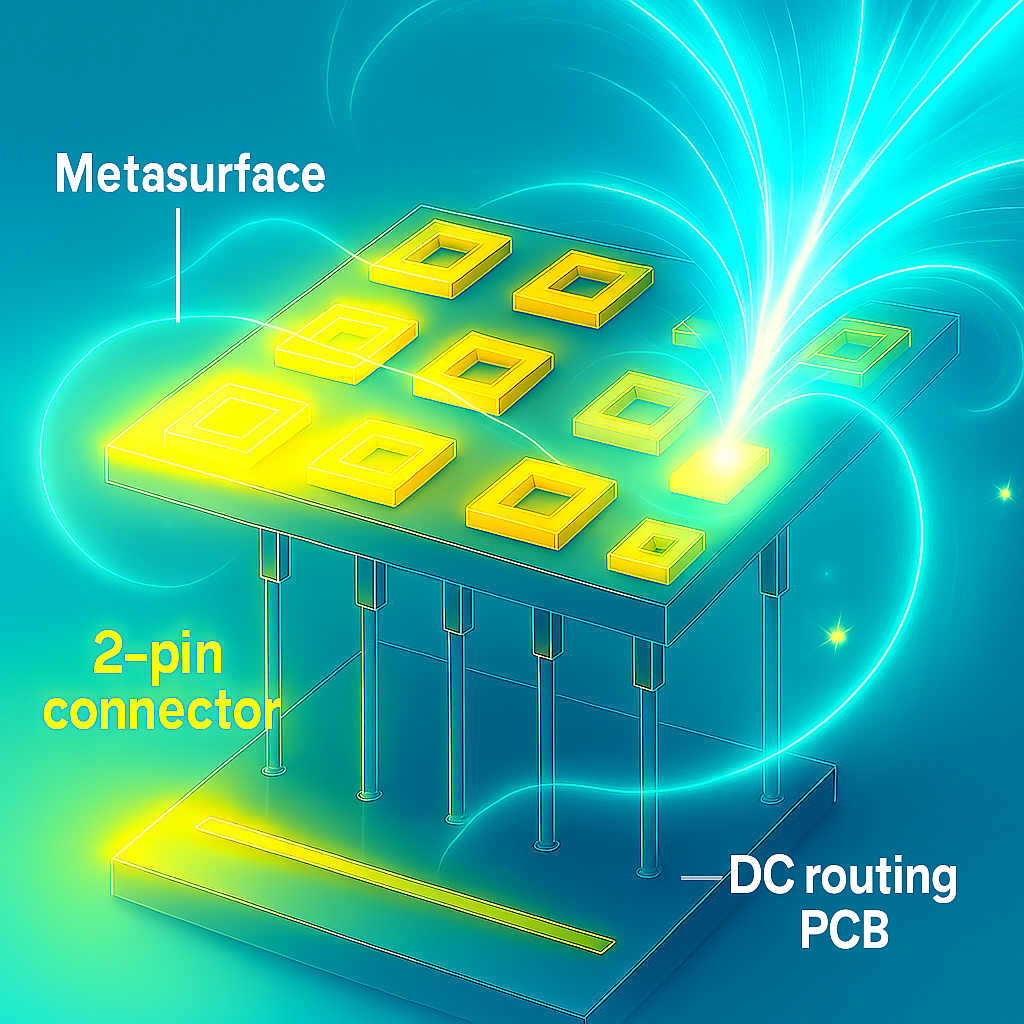}
	\caption{Conceptual diagram of RF-DC separation architecture.}
	\label{fig:concept}
\end{figure}

Among the various metasurface designs, 2-bit metasurfaces stand out due to their balance between performance and complexity. By supporting four discrete phase states, 2-bit metasurfaces achieve precise phase control, which makes them particularly suitable for applications requiring high gain and low sidelobe levels, such as long-distance communication and radar detection. 

Despite their advantages, 2-bit metasurfaces face significant challenges in achieving large-scale array scalability. The primary limitation arises from the complexity of the direct current (DC) control networks required to individually address each element. As the array size increases, the dense routing of DC control lines becomes a bottleneck, leading to increased fabrication complexity, higher losses, and reduced reliability, especially in high frequency band. This issue severely restricts the ability to extend 2-bit metasurface arrays to large-scale configurations, limiting their potential in applications requiring expansive apertures, such as high-gain antennas for satellite communications or wide-area radar systems. Therefore, enabling the large-scale extension of 2-bit metasurfaces has long been a challenging issue and constitutes a central focus of our research efforts.

To address the aforementioned issues, in this work, we propose a novel 2-bit metasurface operating in the Ku-band. To address the critical challenge of array scalability, we introduce a pioneering RF-DC separation architecture. This innovative design decouples the radio frequency (RF) metasurface array from the DC control circuitry by integrating them onto separate printed circuit boards (PCBs), interconnected via pin cascading. This approach enables theoretically unlimited two-dimensional array expansion, overcoming the limitations of conventional DC routing. The proposed metasurface and its scalable architecture provide a robust framework for large-scale electromagnetic wave manipulation, offering significant reference value for next-generation wireless systems and radar technologies.

\section{2-bit Element Design}
\subsection{Structure and Theorem of Element}
\begin{figure}[t]

        \centering
        \includegraphics[width=0.99\linewidth]{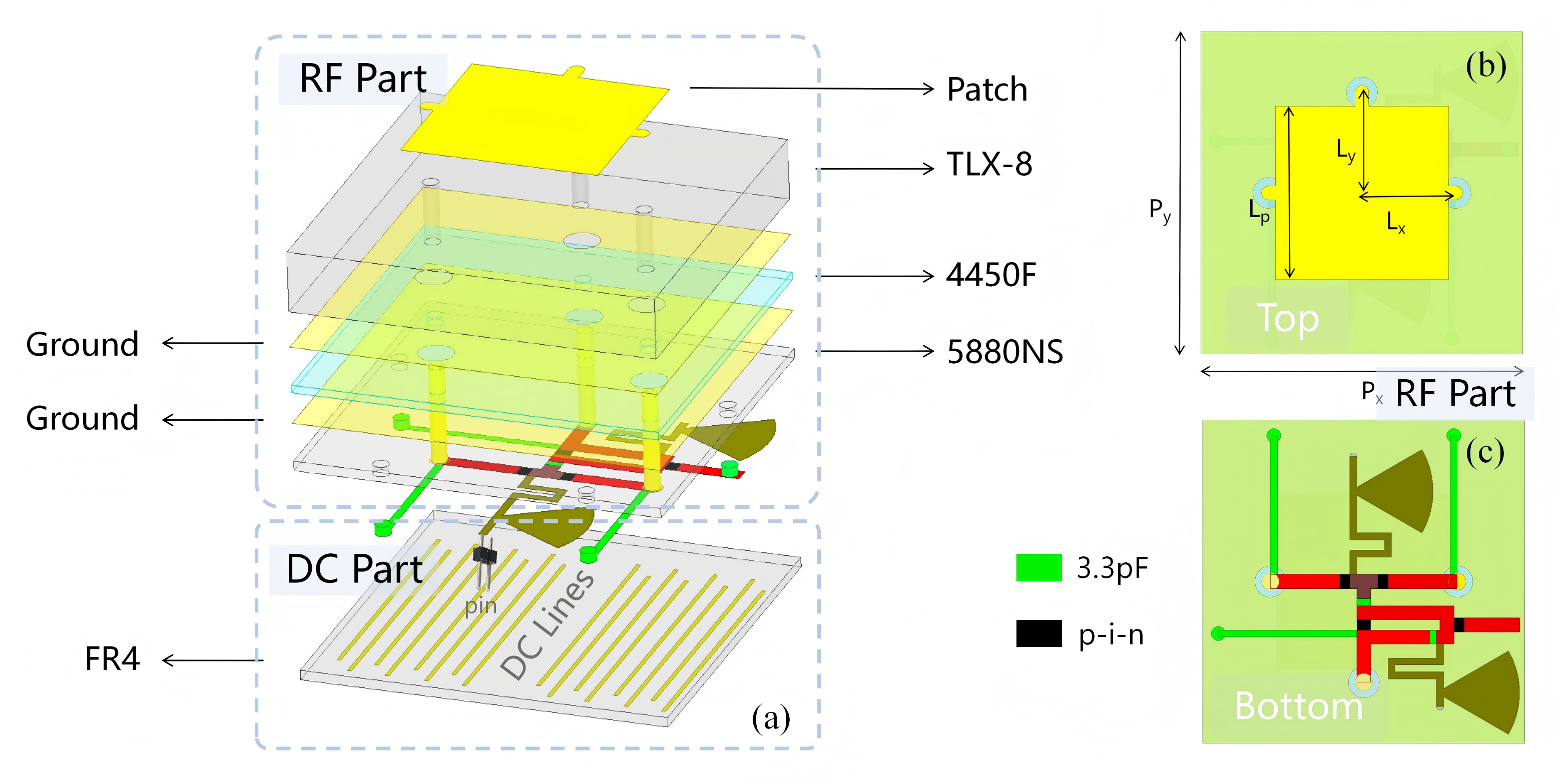}
    \caption{Element structure:(a) RF-DC separation architecture topology; (b) RF part top view; (c) RF part bottom view.}
    \label{fig:overall}
\end{figure}

The proposed 2-bit polarization-conversion element is designed to operate in the Ku-band, with a center frequency of 16.2\,GHz and an effective operating bandwidth ranging from 15\,GHz to 17\,GHz. As depicted in Fig.\ref{fig:overall}, the element consists of an RF part and a DC part: The DC part features a straightforward structure, requiring only routing, and is isolated from the RF part, ensuring no impact on its performance. Thus, the design primarily focuses on the RF part. 

The RF part comprises two layers of dielectric substrates: The top substrate was fabricated using Taconic TLX-8 material, characterized by a relative permittivity of $\varepsilon_r = 2.55$, a loss tangent of $\tan\delta = 0.0018$, and thicknesses of $h_1 = 1.57\,\mathrm{mm}$. The bottom substrate utilizes Rogers RT/duroid 5880, with $\varepsilon_r = 2.2$, $\tan\delta = 0.0009$, and $h_2 = 0.254\,\mathrm{mm}$. The two substrates are laminated together using Rogers 4450F prepreg, which has a relative permittivity of $\varepsilon_r = 3.52$, a loss tangent of $\tan\delta = 0.004$, and a thickness of $h = 0.2\,\mathrm{mm}$.

\begin{table}[t]
\centering
\caption{Key Parameters (Unit: mm)}
\normalsize
\setlength{\tabcolsep}{3.2pt} 
\begin{tabular}{ccccccccc}
\toprule
Parameter & $P_x$ & $P_y$ & $L_p$ & $L_x$ & $L_y$ & $l_1$ & $l_2$ & $l_3$ \\
Value     & 9.3   & 9.3   & 5.0   & 2.7   & 2.9  & 0.4   & 2.8   & 1.6   \\

\bottomrule
\end{tabular}
\label{tab:patch_parameters}
\end{table}

The top layer of the element comprises a metallic patch with three feed points distributed in distinct horizontal directions. This arrangement allows the structure to effectively receive incident electromagnetic waves of both X- and Y-polarizations, the specific design parameters of the patch are shown in Table \ref{tab:patch_parameters}. The received wave is then converted into surface current and transferred to the bottom phase shifting layer through the through-hole. The phase shifting layer incorporates a $90^\circ$ phase shifter and a $180^\circ$ current reversal structure. By selectively switching PIN diodes (MADP-000907-14020, the equivalent model of the diode on and off state is shown in Fig.~\ref{fig:phaseshifter_structure}(c)), the structure achieves discrete reflection phases of $0^\circ$, $90^\circ$, $180^\circ$, and $270^\circ$, thereby realizing a 2-bit coding functionality. Notably, the input and output currents associated with the patch exhibit orthogonal polarization directions, resulting in the desired polarization conversion effect.

\subsection{$90^\circ$ Phase Shifter Design}
This section presents the design of the proposed 90$^\circ$ electrically tunable digital phase shifter, elaborates on its operating principle, and evaluates its performance characteristics.

\begin{figure}[t]
	\centering
	\includegraphics[width=0.98\linewidth]{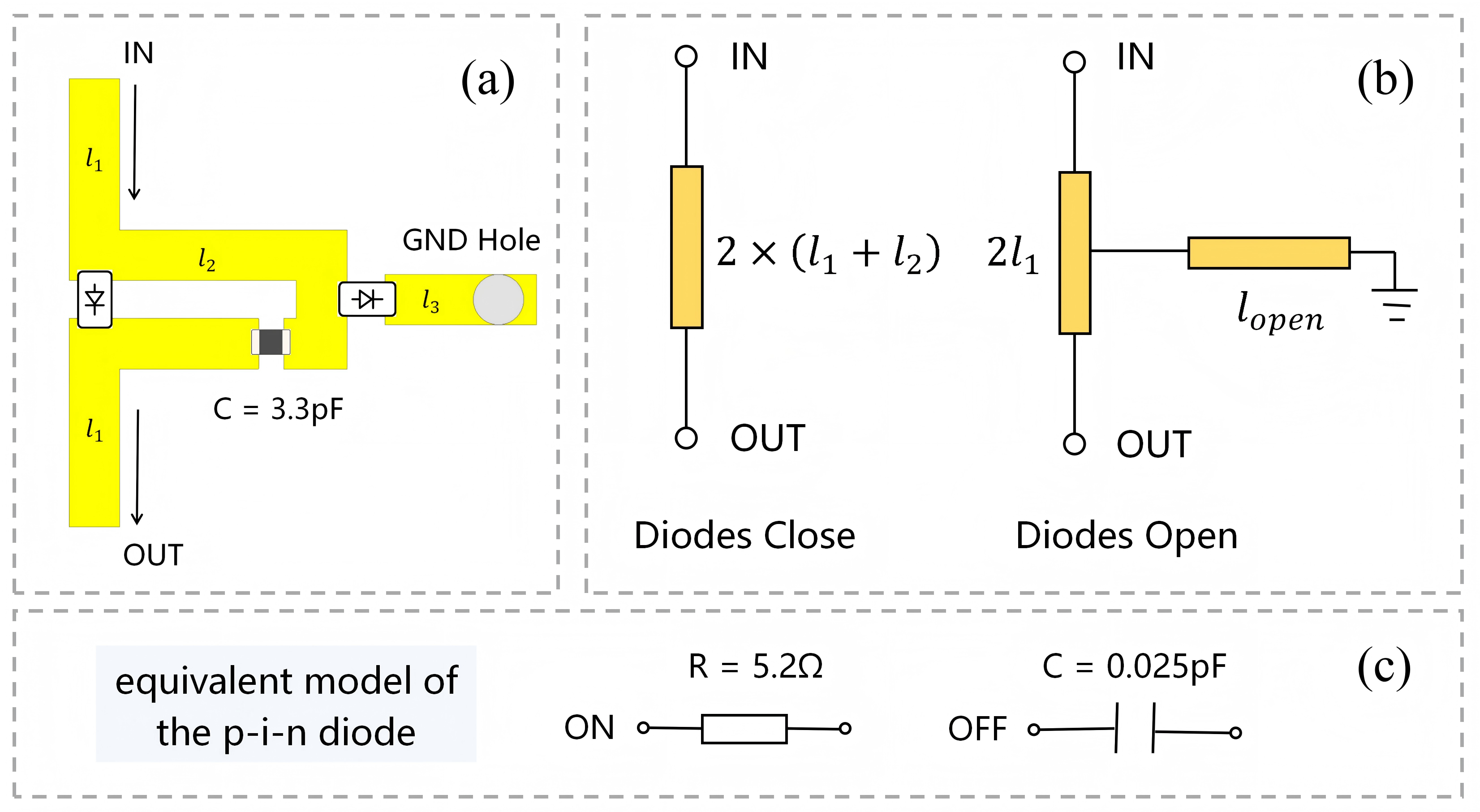}
	\caption{$90^\circ$ phase shifter schematic: (a) structure; (b) two states; (c) diode equivalent circuit.}
	\label{fig:phaseshifter_structure}
\end{figure}

The structural layout of the phase shifter is illustrated in Fig.~\ref{fig:phaseshifter_structure}(a). The design comprises a bent microstrip transmission line, two PIN diodes, a DC-blocking capacitor, and a grounded via. The signal input and output ports are located at the two ends of the microstrip line. The central bent section of the microstrip consists of two symmetric arms. On one side, a PIN diode is connected such that one terminal is soldered to the microstrip line while the other is grounded via a metal through-hole. The length of this arm is approximately $2 \times l_2$. On the opposite side, a second PIN diode is directly bridged between two segments of microstrip lines, each with length $l_1$.

The phase shifting behavior is determined by the biasing states of the PIN diodes, as depicted in the equivalent circuits in Fig.~\ref{fig:phaseshifter_structure}(b). When both diodes are in the reverse-biased (OFF) state, the signal propagates along the full length of the bent microstrip, yielding an effective electrical path of approximately $2 \times (l_1 + l_2)$. In contrast, when both diodes are forward-biased (ON), the structure behaves as follows: the diode on the longer arm shorts the transmission line to ground, forming a short-circuited stub with length $l_3$. The resulting current distribution along this stub exhibits a standing wave pattern. By optimizing the length $l_3$, the stub presents an effective open-circuit impedance at its junction with the main line due to quarter-wavelength resonance, thereby isolating that branch. Consequently, the signal is forced to propagate through the shorter arm only, yielding an effective electrical length of $2 \times l_1$. Due to the difference in electrical lengths between the two switching states, the transmitted signal undergoes distinct phase delays. In this design, the longer path is chosen such that $2 \times l_2$ approximately corresponds to a quarter of the guided wavelength at the center frequency, resulting in a differential phase shift of 90$^\circ$ between the two states. These two discrete operating states are herein referred to as the ``0$^\circ$ state'' and the ``90$^\circ$ state,'' respectively. The key parameters of the phase shifter are shown in Table~\ref{tab:patch_parameters}.

\begin{figure}[t]
	\centering
	\includegraphics[width=1\linewidth]{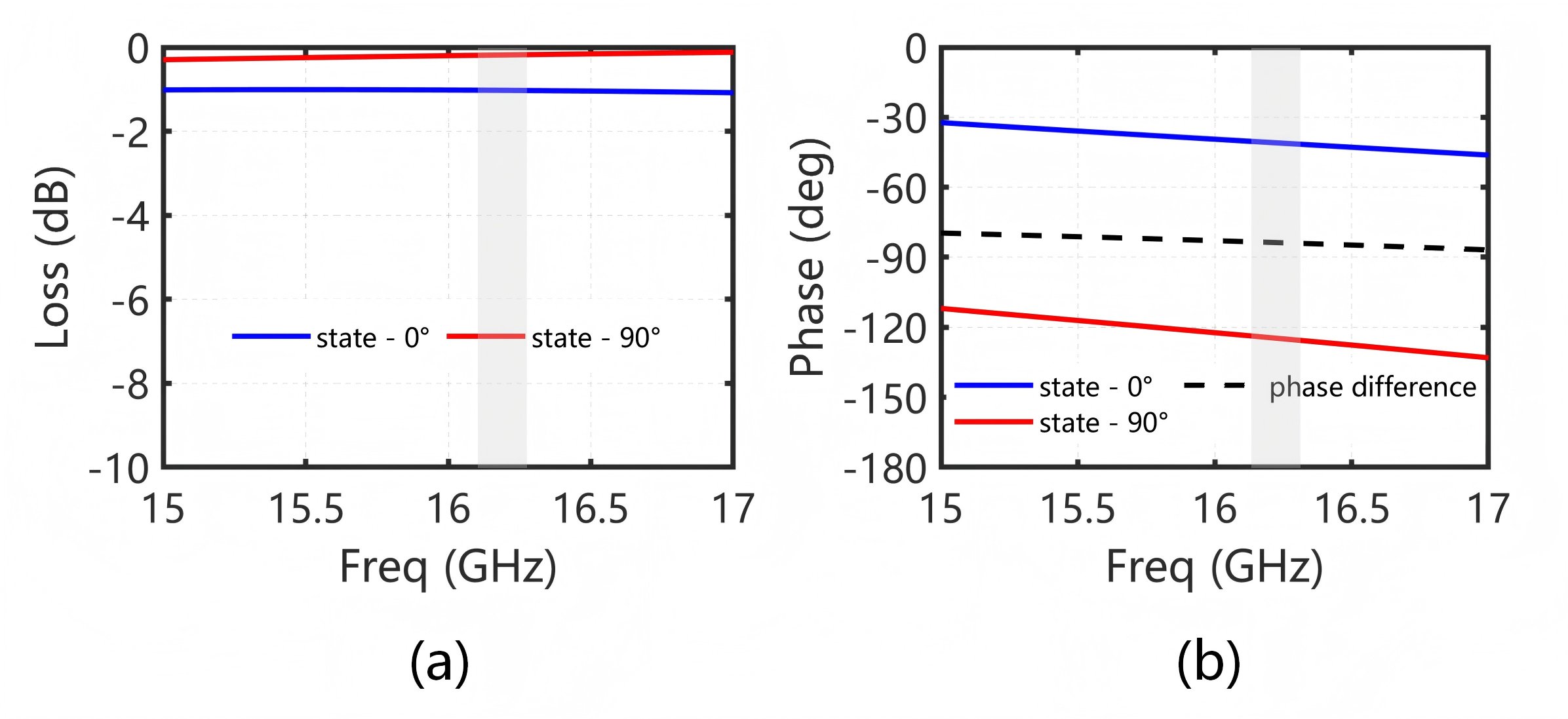}
	\caption{$90^\circ$ phase shifter performance.}
	\label{fig:phaseshifter_simulation}
\end{figure}

The electromagnetic performance of the proposed phase shifter was evaluated using $Ansys\ HFSS$, a full-wave 3D electromagnetic simulation tool. Wave ports were employed at the input and output terminals to excite and measure the device response. The simulation results are presented in Fig.~\ref{fig:phaseshifter_simulation}. At the target operating frequency of 16.2~GHz, the insertion losses for the 0$^\circ$ and 90$^\circ$ phase states are approximately 0.18~dB and 1.02~dB, respectively. These low insertion loss values indicate efficient transmission characteristics in both switching states. Moreover, a clear phase difference of approximately 90$^\circ$ is observed between the two states, demonstrating the successful realization of the desired phase shifting functionality.

In addition to the performance at the center frequency, the phase shifter exhibits broadband characteristics. Across a wide frequency range, the phase difference remains close to 90$^\circ$, confirming the robustness and frequency-insensitive nature of the design. This broadband behavior makes the phase shifter particularly suitable for applications requiring consistent phase performance over a wide operational bandwidth.

\subsection{Element Performance}
After passing through the 90$^\circ$ phase shifter, the signal current enters the 180$^\circ$ phase shift region. This structure is implemented using a symmetric pair of microstrip transmission lines and two PIN diodes, forming a classic current-reversal architecture. The two PIN diodes are operated in complementary switching states—when one diode is turned on, the other is turned off. This configuration forces the current to propagate along opposite paths depending on the diode states. As a result, the excitation direction of the surface current on the patch is reversed, generating electric fields with a 180$^\circ$ phase difference.

\begin{figure}[t]
	\centering
	\includegraphics[width=0.99\linewidth]{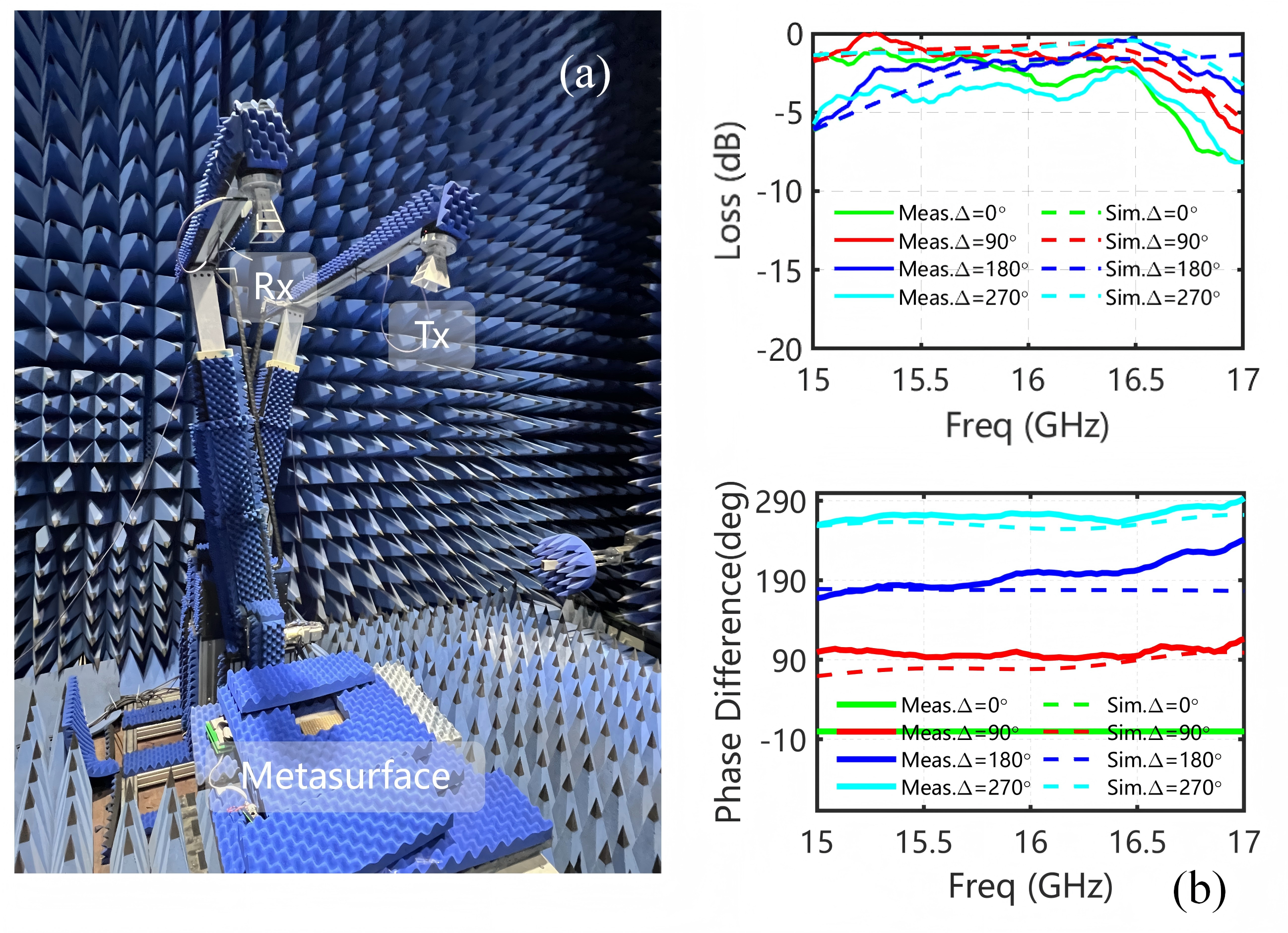}
	\caption{Element measurement: (a) test environment; (b) simulation and measurement results.}
	\label{fig:element_simulation}
\end{figure}

As illustrated in Fig.~\ref{fig:element_simulation}, at the center frequency of 16.2~GHz, the simulated insertion losses for the four phase states are 1.59~dB, 0.62~dB, 1.62~dB, and 0.67~dB,the reflection phase are $-155^\circ, -73^\circ, 28^\circ, 109^\circ$, respectively. These results demonstrate that the proposed element achieves stable four-state phase control over a wide frequency band. Moreover, the four phase response curves remain nearly parallel across the operating band, with an approximate and consistent spacing of 90$^\circ$ between adjacent states. Furthermore, the measured phase responses are in good agreement with the simulated results, confirming the accuracy of the design. The amplitude responses also generally align well with simulation, indicating stable performance. A slight frequency shift toward the lower end of the spectrum is noted, which is presumed to be caused by fabrication tolerances and processing deviations. Despite this minor shift, the measured results validate the effectiveness of the proposed 2-bit element in achieving broadband and low-loss phase modulation.

In addition, an oblique incidence analysis was carried out. Simulation results indicate that under a $30^\circ$ oblique incidence, the phase variation remains within $20^\circ$.

\section{Metasurface Fabrication and Measurement}

\subsection{Fabrication of Infinitely Scalable Structure}

\begin{figure}[t]
	\centering
	\includegraphics[width=0.85\linewidth]{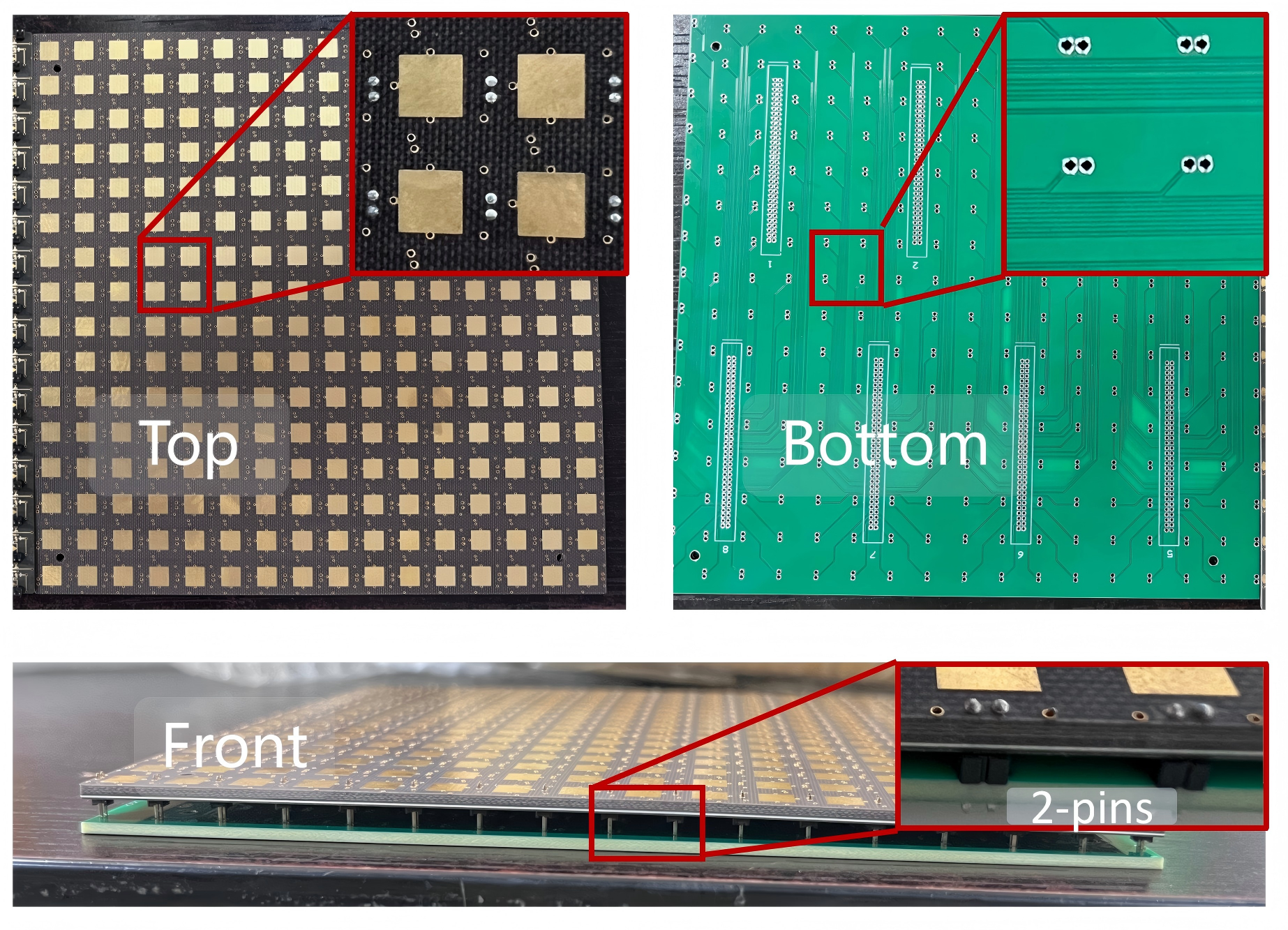}
	\caption{Physical picture of the separation architecture metasurface.}
	\label{fig:demo_photo}
\end{figure}

Based on the proposed separation architecture, four 16$\times$16 infinitely extendable prototype arrays were fabricated. The PCB layouts of the metasurface layer and the DC control layer are independently fabricated, as illustrated in Fig.~\ref{fig:demo_photo}. To ensure seamless interconnection between the two layers, each element reserves two dedicated via positions located identically in both PCBs. The pins connect the feeding points of the metasurface to the corresponding positions on the DC routing PCB, which is subsequently interfaced with the control board and the FPGA-based voltage board. This hierarchical connection scheme ensures precise and reliable current allocation to each element, enabling stable phase control across the entire metasurface.

\subsection{Measurement of $4\times16\times16$ Metasurface}

\begin{figure}[t]
	\centering
	\includegraphics[width=0.95\linewidth]{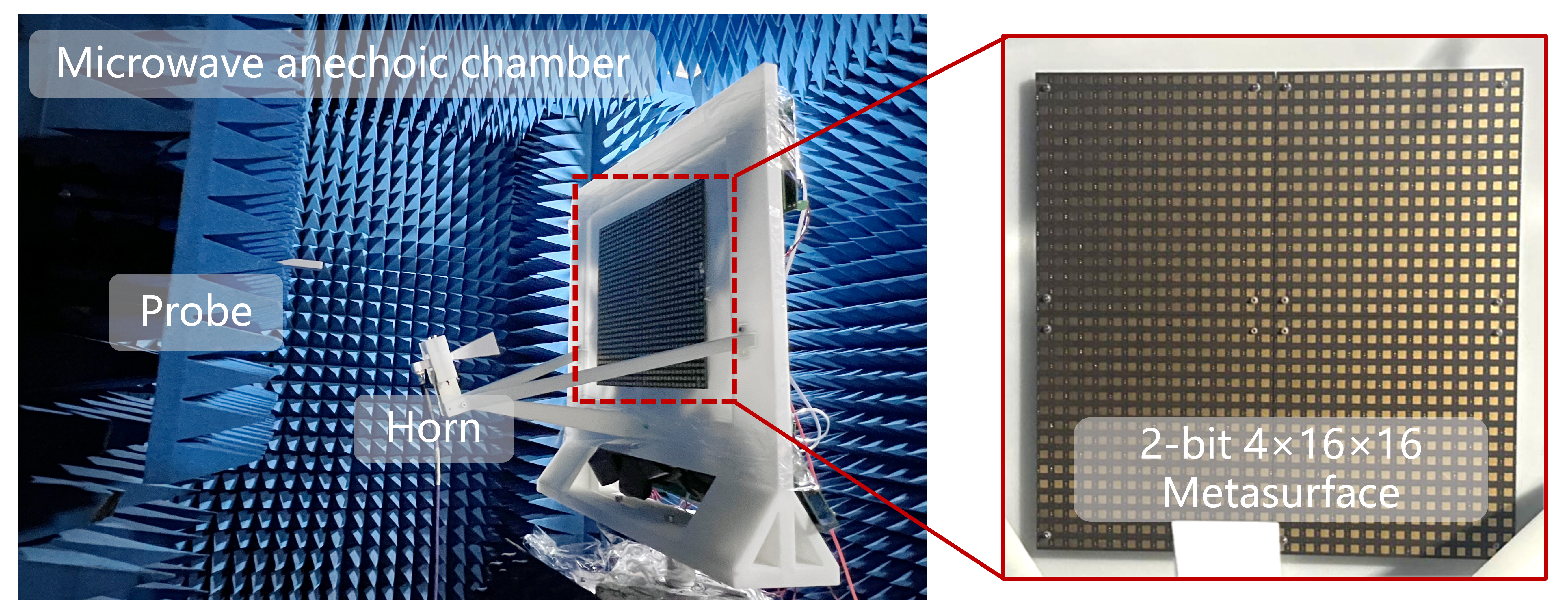}
	\caption{Measurement environment of separation architecture meatsurface.}
	\label{fig:darkroom_setup}
\end{figure}

Subsequently, the assembled prototypes were tested in a planar far-field measurement chamber. The measurement setup is shown in Fig.~\ref{fig:darkroom_setup}. The horn antenna still operates under a 20° oblique incidence, with the focal ratio fixed at 0.9.

\begin{figure}[t]
	\centering
	\includegraphics[width=0.99\linewidth]{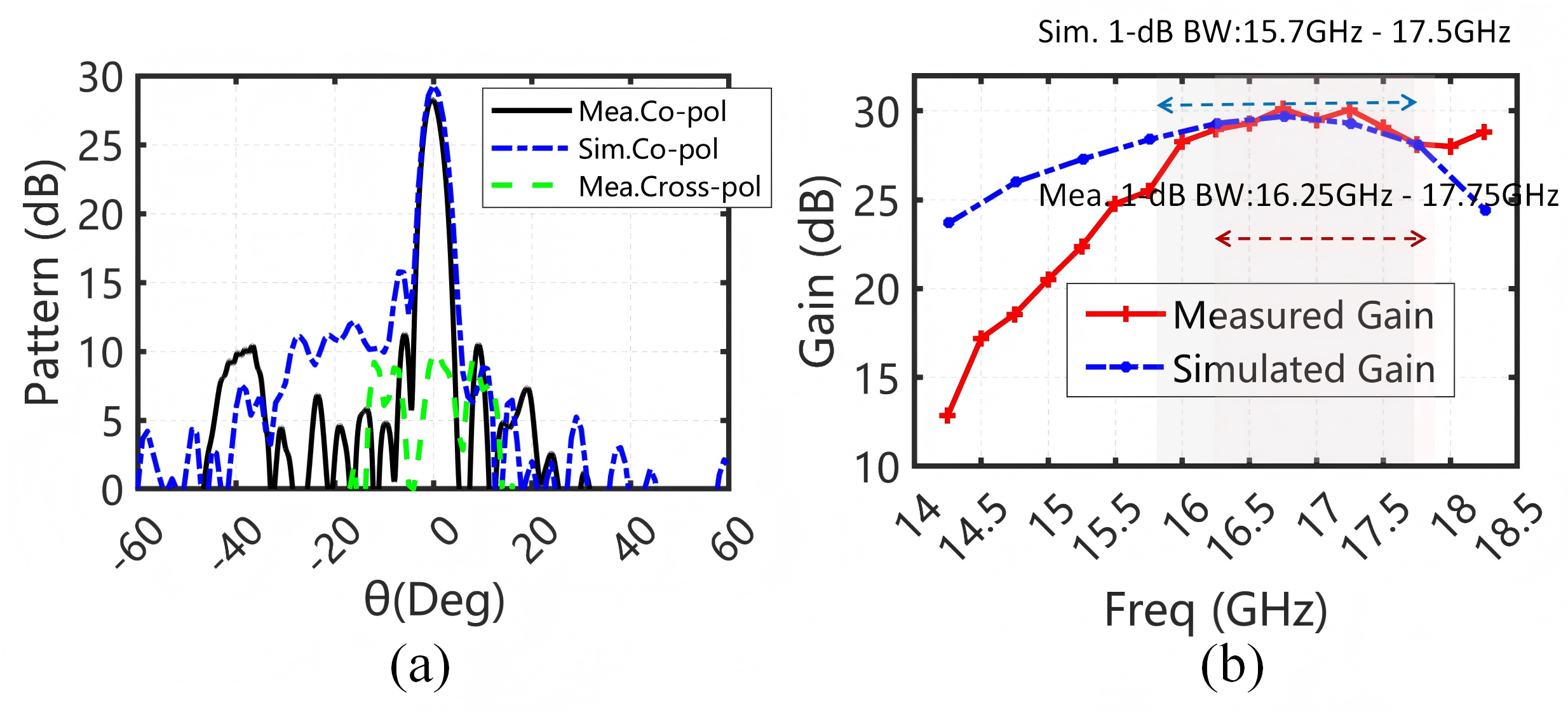}
	\caption{(a) $0^\circ$ beam radiation pattern performance comparision; (b) gain versus frequency curve.}
	\label{fig:3232_0deg}
\end{figure}

First, the measured results of the normal beam were evaluated. As shown in the Fig.\ref{fig:3232_0deg}(a), the measured radiation pattern is generally consistent with the $CST$ full-wave simulation results. According to the measurement results, the metasurface achieves a maximum boresight gain of 28.3\,dB, corresponding to an aperture efficiency of 21.02\%, which is approximately 1dB lower than the simulated value. The sidelobes can be suppressed to below $-18$\,dB, even better than the simulation results. The measured cross-polarization level is approximately 19dB lower than the co-polarization level, indicating good polarization purity and radiation performance.

\begin{figure}[t]
	\centering
	\includegraphics[width=0.99\linewidth]{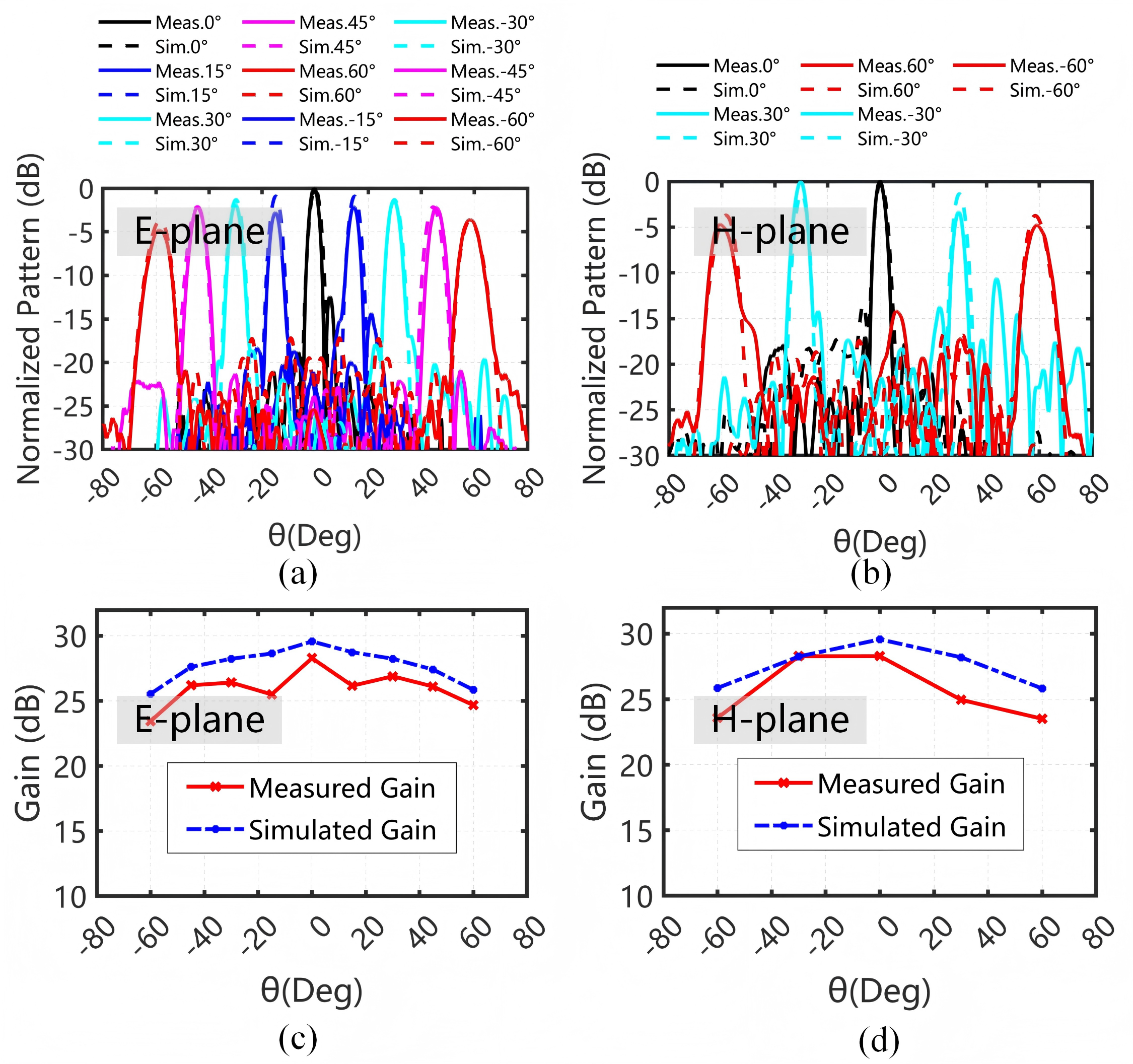}
	\caption{Measurement results: (a) E-plane normalized radiation pattern of scanning: From -60° to 60° in 15° steps; (b) H-plane normalized radiation pattern of scanning: From -60° to 60° in 15° steps; (c) E-plane scanning gain comparision of measurement and simulation; (d) H-plane scanning gain comparision of measurement and simulation.}
	\label{fig:E-plane radiation pattern of horizontal scanning}
\end{figure}

The two-dimensional beam scanning performance of the metasurface was experimentally evaluated. Specifically, beam scanning was performed on the E-plane and H-plane, respectively. As shown in Fig.~\ref{fig:E-plane radiation pattern of horizontal scanning}, the normalized patterns of measurement and simulation are in good agreement. Furthermore, a comparison was made between the simulated and measured gain variation curves with respect to the radiation angle. The measured results are slightly lower than the simulated ones, which can be attributed to fabrication tolerances, environmental influences during testing, and limitations in measurement equipment accuracy. According to the measurement results, the metasurface achieves high gain and low SLL performance. Moreover, two-dimensional beam scanning up to $\pm 60^\circ$ was successfully demonstrated, with a minimum scanning loss of only 3.61\,dB.

Finally, the variation of the array boresight gain with frequency was measured and compared with the corresponding simulation results. From the test results in Fig.\ref{fig:3232_0deg}(b), it can be observed that the overall measured gain curve shifts slightly towards higher frequency compared to the simulation. This discrepancy is likely due to fabrication tolerances associated with the new structural design. Nevertheless, the results demonstrate a favorable gain bandwidth performance. The simulated 1-dB gain bandwidth ranges from 15.7\,GHz to 17.5\,GHz, while the measured 1-dB gain bandwidth spans from 16.25\,GHz to 17.75\,GHz (defined as the frequency interval within 1\,dB of the peak gain), corresponding to a relative bandwidth of 9.26\%.


\begin{table}[t]
\centering
\caption{Comparison of Different 2-bit Metasurface Performances}

\scriptsize
\renewcommand{\arraystretch}{0.95}

\setlength{\tabcolsep}{2.2pt} 
\begin{tabular}{ccccccc}
\toprule
Reference & Frequency & Polarization & Array Size & Measured Gain & Scan Range & Scalable \\
\midrule
\cite{b20} & 2.3 GHz  & LP       & $16 \times 16$ & 21.7 dB  & $\pm50^\circ$ & No  \\
\cite{b15} & 3.5 GHz  & Dual-LP  & $12 \times 12$ & 16.9 dB  & $\pm40^\circ$ & No  \\
\cite{b21} & 3.5 GHz  & Dual-LP  & $12 \times 12$ & 20.2 dB  & $\pm60^\circ$ & No  \\
\cite{b22} & 3.5 GHz  & LP/CP    & $16 \times 16$ & 20.2 dB  & $\pm45^\circ$ & No  \\
\cite{b13} & 5 GHz    & LP       & $16 \times 16$ & 23.5 dB  & $\pm60^\circ$ & No  \\
\cite{b14} & 10.2 GHz & LP       & $12 \times 12$ & 16.7 dB  & $\pm60^\circ$ & No  \\
\cite{b16} & 29 GHz   & LP       & $14 \times 14$ & 19.8 dB  & $\pm60^\circ$ & No  \\
This work  & 16.2 GHz & Dual-LP  & $32 \times 32$ & 28.3 dB  & $\pm60^\circ$ & Yes \\
\bottomrule
\end{tabular}

\label{tab:2bit_metasurface_comparison}
\end{table}

From the test results, it can be seen that the proposed RF-DC separation architecture achieves excellent performance, more importantly it enables seamless scalability of the antenna array. A comparison between the proposed 2-bit metasurface and existing  works is presented in Table~\ref{tab:2bit_metasurface_comparison}, demonstrating clear advantages in terms of polarization, array size, gain, and other performance metrics.

\section{Conclusion}
This letter presents the design of a 2-bit metasurface operating in the Ku band, capable of polarization conversion and supporting dual-polarization control. To address the challenge of routing complexity in large-scale 2-bit high-frequency metasurface arrays, this paper innovatively proposes an RF-DC separation architecture. A $4\times16\times16$ prototype was implemented to verify this approach. The experimental results demonstrate that the proposed structure obtains high antenna performance while enabling seamless two-dimensional scalability. This work provides a promising solution for applications such as long-distance communications and radar detection, where high-gain antenna systems are essential. Moreover, it offers a novel and practical approach for implementing large-scale deployable metasurface arrays in engineering applications.


\end{document}